\documentclass[sigconf]{acmart}
\acmConference[EASE 2025]{The 29th International Conference on Evaluation and Assessment in Software Engineering}{17–20 June, 2025}{Istanbul, Türkiye}

\usepackage{algorithmic}
\usepackage{graphicx}
\usepackage{textcomp}
\usepackage{url}
\usepackage{xcolor}

\usepackage{tabularx}
\usepackage{booktabs}
\usepackage{adjustbox}
\usepackage{framed}
\usepackage{multicol}
\usepackage{multirow}
\usepackage{colortbl} 
\usepackage{multicol}             
\usepackage{geometry}
\usepackage{longtable}
\usepackage{colortbl}
\usepackage{tcolorbox} 

\setcopyright{acmlicensed}
\copyrightyear{2018}
\acmYear{2018}
\acmDOI{XXXXXXX.XXXXXXX}

\acmISBN{978-1-4503-XXXX-X/2018/06}

\def\BibTeX{{\rm B\kern-.05em{\sc i\kern-.025em b}\kern-.08em   T\kern-.1667em\lower.7ex\hbox{E}\kern-.125emX}}

\definecolor{quote}{HTML}{9673A6}

\newcommand{\pquote}[1]{\textcolor[HTML]{9673A6}{\textbf{Participant #1:} }}

\newcommand{\ta}[1]{\textcolor[HTML]{6C8EBF}{\textbf{#1} }}

\newenvironment{zeroindent}
  {\par\setlength{\parindent}{0pt}}
  {\par}

\usepackage[framemethod=TikZ]{mdframed}
\usepackage{xcolor} 

\definecolor{customcolor}{HTML}{6C8EBF}

\newenvironment{fancy}[2][]{
    \mdfsetup{
        skipabove=1pt, 
        innerlinewidth=1pt, innerlinecolor=#2, 
        linewidth=0pt,
        backgroundcolor=#2!20 
    }
    \begin{mdframed}}
    {\end{mdframed}}

\newcommand{\cristysbox}[1]{
\vspace{5pt}
\begin{fancy}[]{customcolor} 
\noindent\parbox{0.96\linewidth}
{\vspace{1px}
#1
} 
\end{fancy}
}

\newenvironment{coloredframe}[2][]{
    \mdfsetup{
        skipabove=2pt, 
        hidealllines=true, leftline=true,      
        innerlinewidth=2pt, innerlinecolor=#2, 
        linewidth=0pt,
        backgroundcolor=#2!10
    }
    \begin{mdframed}}
    {\end{mdframed}}

\newcommand{\dialoguegpt}[2]{
    \vspace{5px}
    \begin{coloredframe}{#1}

    \begin{zeroindent} #2 \end{zeroindent}
    \vspace{1px}
    \end{coloredframe}
    \vspace{5px}
}

\begin{document}

\title{Emotional Strain and Frustration in LLM Interactions in Software Engineering}


\author{Cristina Martinez Montes}
\email{montes@chalmers.se}
\orcid{0000-0003-1150-6931}
\affiliation{%
  \institution{Chalmers University of Technology and University of Gothenburg}
  \city{Gothenburg}
  \country{Sweden}
}

\author{Ranim Khojah}
\email{khojah@chalmers.se}
\orcid{0000-0002-1090-3153}
\affiliation{%
  \institution{Chalmers University of Technology and University of Gothenburg}
  \city{Gothenburg}
  \country{Sweden}
}


\renewcommand{\shortauthors}{Montes et al.}

\begin{abstract}
Large Language Models (LLMs) are increasingly integrated into various daily tasks in Software Engineering such as coding and requirement elicitation. Despite their various capabilities and constant use, some interactions can lead to unexpected challenges (e.g. hallucinations or verbose answers) and, in turn, cause emotions that develop into frustration. Frustration can negatively impact engineers' productivity and well-being if they escalate into stress and burnout. In this paper, we assess the impact of LLM interactions on software engineers' emotional responses, specifically strains, and identify common causes of frustration when interacting with LLMs at work.
Based on 62 survey responses from software engineers in industry and academia across various companies and universities, we found that a majority of our respondents experience frustrations or other related emotions regardless of the nature of their work. Additionally, our results showed that frustration mainly stemmed from issues with correctness and less critical issues such as adaptability to context or specific format. While such issues may not cause frustration in general, artefacts that do not follow certain preferences, standards, or best practices can make the output unusable without extensive modification, causing frustration over time.
In addition to the frustration triggers, our study offers guidelines to improve the software engineers' experience, aiming to minimise long-term consequences on mental health.







\end{abstract}

\keywords{Software Engineering, Large Language Models (LLMs), Frustration, Emotions}

\maketitle

\section{Introduction}
Software Engineering (SE) comes with many challenges, from fixing bugs to dealing with changing requirements. Recently, Large Language Models (LLMs) and LLM-powered chatbots like ChatGPT and GitHub Copilot have been used by software engineers to assist them in performing various tasks including code generation, and quality assurance \citep{asare2023github, lubos2024leveraging}. Current research focuses on understanding how practitioners aim to increase their productivity and make their work process more efficient by targeting LLMs to automate the generation of software artifacts or receive guidance on how to solve certain problems \citep{khojah2024beyond, storey2016disrupting}.

Challenges and limitations of LLMs hinder their effectiveness, such as unhelpful responses, which can lead to frustration among engineers \citep{weisz2022better}. This frustration contributes to techno-stress, affecting their workflow, well-being, and productivity \citep{tarafdar2007impact, muller2023can}.
While frustrations have gotten little attention in LLM research for software engineering, we argue that understanding the causes of frustrations when using LLMs for software-related tasks is the first step to minimising them and thus improving the productivity and well-being of practitioners in the SE industry and academia. Moreover, revealing such triggers helps the designers of LLM-powered tools (e.g., AI Chatbots) improve the user experience and evolution of such tools.

This exploratory study aims to empirically investigate the causes of frustrations in software engineers' interactions with LLMs (and LLM-powered chatbots) and propose strategies for improvement. We focus on the following research questions: 

\noindent \emph{RQ1: What are the triggers or sources of frustration of software engineers when using LLMs?} 

Our study presents four main categories that can cause frustrating emotions for the software engineer. We found that the main cause of frustration is when the software engineers receive an unhelpful or incorrect answer, followed by misunderstanding the intention, failing to meet personal preferences, and other limitations of LLMs. We argue that these categories are specific to SE since it is a domain that requires precise, context-aware, and technically accurate responses that can be directly applied, like using LLMs for code generation. 

\noindent \emph{RQ2: How can the frustrating experience impact motivation?}

We report that while frustration from unmet expectations can momentarily impact motivation, software engineers typically remain engaged in tasks despite these frustrating interactions with LLMs. This suggests that such interactions are not usually disruptive enough to prevent task completion.

\noindent \emph{RQ3: How can the user experience be improved to reduce frustration for software engineers?} 

We provide recommendations for improvements based on software engineers' expectations and lived experiences. Engineers offer practical, grounded recommendations that reflect user needs and expectations since they are the primary stakeholders directly interacting with LLMs in real-world contexts and can guide chatbot designers in enhancing design and usability.
 
In addition, we suggest managers provide training and raise awareness among software engineers in order to manage and minimise frustrations.


\section{Background and Related Work}

This section presents the conceptual framework of this study, as well as, previous work done on the topic and the the research gap addressed.

\subsection{Large Language Models in Software Engineering} 

In SE, recent research has shown that LLMs have the potential to support practitioners in a variety of tasks including, but not limited to implementation, testing, requirements engineering \cite{dakhel2023github, wang2024software, norheim2024challenges, khojah2024beyond}.
Moreover, despite the challenges LLMs impose on academia (e.g., bias and hallucination), they provide many opportunities for researchers and educators as assistants with creating study guides and academic writing \cite{meyer2023chatgpt}.

Despite this extensive use and the time saved when automating tasks, integrating LLMs into SE practices can lead to user frustration, which can be understood as the emotional state experienced by a person when they are prevented or hindered from obtaining something they have been led to expect \cite{apa_dictionary_frustration}.

Researchers have noted that while LLMs assist software practitioners, they may sometimes generate errors, causing them to spend additional time solving errors or seeking clarification. This can also trigger negative emotions, such as frustration \citep{weisz2022better}.

Our study addresses a gap in the literature by exploring whether the above findings also apply to LLM interactions for SE tasks, with a focus on frustration.

\subsection{Emotions Involved When Using Technology}
\begin{figure}
    \centering
    \includegraphics[width=1\linewidth]{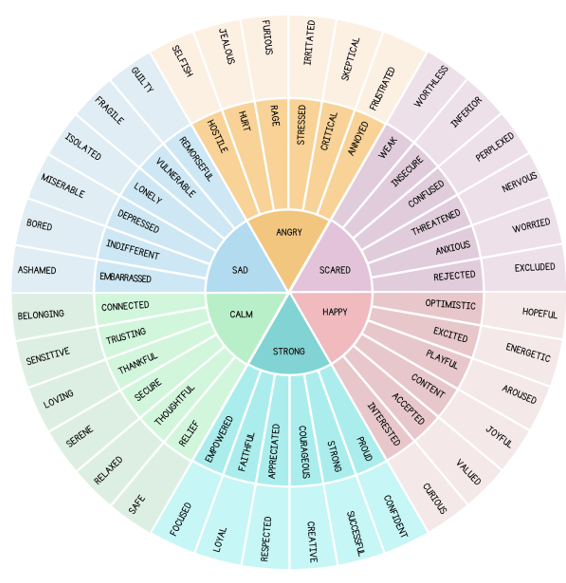}
    \caption{Willcox's Emotions Wheel \citep{neff2024feelings}}
    \Description[]{}
    \label{fig:wheel}
\end{figure}

In this study, we considered the definition by the American Psychological Association~\citep{apa_dictionary_emotion}, Emotion is \textit{``a complex reaction pattern, involving experiential, behavioural, and physiological elements, by which an individual attempts to deal with a personally significant matter or event''}. We acknowledge the diversity of the conceptual definitions of emotions, we selected this definition for its scope, which integrates experiential, behavioural, and physiological dimensions. 
The Emotions Wheel, developed by Gloria Willcox~\citep{willcox1982feeling}, is a psychological tool designed to facilitate identifying and expressing emotions. It is widely used in therapeutic settings and personal development to help individuals articulate their emotional states more precisely. The tool organises emotions into a concentric structure, with six core emotions—happy, sad, angry, scared, strong, and calm—situated at the center of the wheel (See Figure\ref{fig:wheel}). As one moves outward from the core, these primary emotions subdivide into more specific and nuanced feelings, allowing for a more granular understanding of emotional experiences.

In this study, we used the adaptation done by \citep{neff_emotional_2023} to explore how individuals experience their interaction with LLMs.

While tools like the Emotions Wheel help categorise and articulate emotional experiences, technology-specific contexts introduce unique emotional challenges. These include emotional strain, which refers to adverse reactions and feelings triggered by stressors \citep{chang2007emotional}, and technostrain and its components, particularly relevant for understanding interactions with LLMs.

Salanova~\citep{salanova2014technostress} defines Technostrain as a negative psychological experience characterised by i) heightened anxiety and fatigue (affective dimension), ii) scepticism (attitudinal dimension), and iii) a sense of inefficacy (cognitive dimension) associated with technology use. Aligned with this, Muller et al.~\citep{muller2023can} researched techno-frustration, a component of technostrain, which refers to the psychological strain caused by the disorganised or inefficient use of Information and Communications Technology (ICT). Techno-frustration describes the experience of feeling discouraged, uncertain, stressed, confused, and upset as a result of using ICT. Such psychological strain can lead to decreased job satisfaction or an increased risk of burnout~\citep{srivastava2015technostress}. Muller et al.'s work is the closest to our study; however, we focus on a specific interaction, LLM-user interaction, which is still underexplored despite the rapid integration of LLMs in SE.

Regarding research on how technology impacts users' emotions, studies have focused on the psychological effects of prolonged technology use, recognising that digital tools influence productivity and mental and emotional states \citep{calvo2014positive, mark2015focused}. 
Wester et al. \citep{wester2024ai} found that incorrect outputs, which reject the user's request, can lead to frustration and greatly diminish their perception of the LLM's usefulness, appropriateness, and relevance.

This reflects an effort to explore the balance between the benefits of technology and the potential strain it places on users, particularly in high-demand environments like SE. Examples of those efforts are in the form of interventions \citep{montes2024qualifying}, \citep{penzenstadler2022take} or looking for the causes \citep{wong2023mental}.

Furthermore, users' expectations when using technology, particularly LLMs, are crucial to their frustration. Studies have shown that prior experiences with technology can shape users' expectations, influencing their perceptions of performance and trust~\citep{ferreri2023identifying}. When technology does not meet users' expectations—whether based on past interactions or external portrayals—frustration can escalate, hindering effective task completion and satisfaction ~\citep{srivastava2015technostress}. By examining these responses, we aim to inform LLM design improvements that enhance productivity and reduce technostress.

\subsection{Emotions in Software Engineering Tasks}

Emotions related to software engineering tasks have been widely researched. Several studies have identified a wide range of emotions, from anger and frustration to joy and satisfaction, as software engineers perform their tasks and communication channels within the development context \cite{foster2012exploring, mantyla2014time, graziotin2014software, marcos2020applying, mantyla2016mining, russell1991culture, Girardi2022}. Sánchez-Gordón's \cite{sanchez2019taking} literature review further analysed the diversity of emotions developers face, identifying 40 discrete emotions, the most frequent being anger, fear, disgust, sadness, joy, love, and happiness. Various situational and contextual factors shape this rich diversity of emotions.
Additionally, the impact of affective states has been related to performance and code quality. For example, Graziotin et al. \cite{graziotin2013happy} found that positive emotions, such as happiness, are closely linked to improved performance and productivity. Conversely, negative emotions, such as frustration and unhappiness, can reduce motivation, hinder task completion, and increase the likelihood of turnover \cite{graziotin2018happens}. Furthermore, studies have found that developers report higher productivity when in a state of flow and often experience frustration due to being stuck, technical difficulties or unfulfilled information needs \cite{muller2015stuck, girardi2020recognizing}.

Moreover, specific triggers of negative emotions have been extensively studied. For example, unhappiness, Graziotin et al. \cite{graziotin2017unhappiness} found that everyday sources of unhappiness are time pressure, bad code quality, repetitive tasks, and inadequate decision-making. Regarding frustration, Ford and Parnin \cite{ford2015exploring} identified program comprehension challenges, poor tooling, and fear of failure as common causes. 

Despite significant research on emotions associated with traditional software engineering tasks, there is still a gap in investigating the emotional impact of interactions with emerging tools like Large Language Models (LLMs). This gap leaves unexplored how emotions such as frustration evolve in response to LLM use, which has critical implications for improving their design and effectiveness in supporting software engineering tasks.

\section{Methodology}
\begin{figure}
    \centering
    \includegraphics[width=1\linewidth]{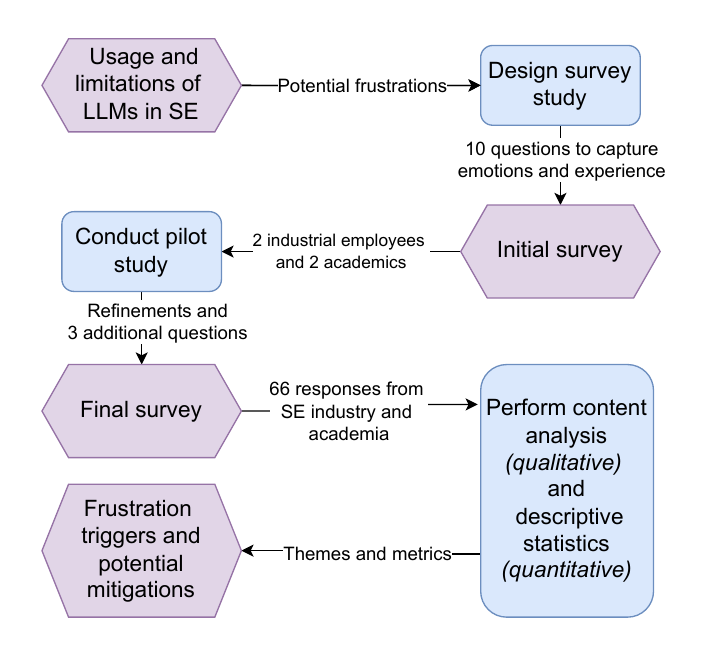}
    \caption{The exploratory study process that we followed to design our survey and analyse the responses qualitatively and quantitively.}
    \Description[study process]{The exploratory study process that we followed to design our survey and analyse the responses qualitatively and quantitively.}
    \label{fig:method}
\end{figure}

Our study aims to understand software engineers' frustrations and emotions when interacting with LLMs, identifying causes and potential solutions. This study implements an exploratory design to gather initial insights and detect interaction patterns~\citep{stebbins2001exploratory}. We used surveys following Stol and Fitzgerald guidelines~\citep{stol2020guidelines} since they are particularly effective for exploratory studies aiming to generalise findings across a population. We collected qualitative and quantitative data through open and closed questions, covering specific and broader aspects of LLM usage. Figure~\ref{fig:method} illustrates our methodology.

\subsection{Target Population}

We surveyed software engineers in academia and industry to get a broader exploration of how LLMs are used in software engineering tasks. While industry engineers provided information about practical applications such as coding, debugging, and meeting production demands, academics, on the other hand, added LLM usage in research-related tasks such as programming, testing and general research on software processes and quality. Despite differences in their environments, both groups share core practices and challenges, such as dealing with tool limitations and managing cognitive load. Including both populations ensured a comprehensive analysis of how LLMs impact software engineering workflows, making our findings broadly relevant to diverse real-world and research contexts.

\subsection{Data Collection}
We used a questionnaire as our main data collection method. We designed the survey questions based on previous research about the usage of LLMs in SE and how it impacts their experience and productivity \citep{khojah2024beyond} as well as general limitations of LLMs \citep{hadi2023survey}. 

After a pilot study with two PhD students (academia) and two software engineers (industry), we added more questions and refined the survey (available in our Zenodo package \citep{zenodo}). The survey started asking for consent to participate and included (i) four demographic questions and queries about participants' experience and familiarity with LLMs, (ii) two open-ended and one closed-ended question about their usage of LLMs during their work and their expectations from these LLMs, (iii) four open-ended questions and one closed-ended question about the participants' emotions when certain expectations are not met and about frustrations specifically that the participants experienced when using LLMs, and (iv) five Likert-scales of the level of importance of different LLMs abilities and aspects to minimise frustrating experiences. 

The survey was created on Google Forms and distributed across social platforms (LinkedIn and Facebook). We used stratified random sampling to get software engineers from academia and industry, drawing on Baltes and Ralph's~\citep{baltes2022sampling} work as a framework for our sampling approach. We sent approximately 20 personal invitations to employees across 10 software organisations of different sizes (Startups and large companies) and domains (e.g., automotive and eLearning). We also invited researchers and academics in SE conferences (RE'24 and FSE'24) to allow our sample of software engineers to be diverse in terms of countries and domains. All data was anonymised, we did not ask for personal data or identification. We followed our university's ethical regulations and guidelines.
We believe that it is important that our sample was diverse in terms of areas and domains, particularly that it included software engineering academics. Software engineering practices often overlap between academia and industry, with some differences in priorities and contexts. This diversity enabled us to capture a broader range of perspectives and demonstrate that academics and practitioners share emotions and challenges related to using LLMs in the field.

\subsection{Data Analysis}


We used content analysis with an inductive approach following the steps by Erlingsson and Brysiewicz~\citep{erlingsson2017hands} to analyse the open questions. Both authors carried out the whole data analysis together systematically. The analysis started by reviewing each answer in detail and discussing them to ensure a shared understanding. This allowed us to identify initial patterns and codes. The codes were then categorised based on their similarities and differences ~\citep{erlingsson2017hands}, with the categories refined iteratively to ensure accuracy. Themes emerged organically from the data, reflecting the participants' perspectives and providing meaningful insights. We used the emotion classification and feeling wheel by Willcox~\citep{willcox1982feeling} to categorise and identify the range of emotions expressed by participants. For example, one participant's comment, "I acknowledge that it might give incorrect answers so it is indifferent for me unless it happens often" was coded as Indiferent'. Enabling a deeper understanding of their emotional responses during interactions with LLMs. For the Likert and close questions, we use descriptive statistics and visualisations.  



\section{Results}
This section presents the results of the visualisation of the closed questions and the content analysis of the open questions. We distinguished between academics and practitioners when their results differed, such as in LLM usage. We combined the results when their patterns were similar, like emotional responses or frustration triggers.

\subsection{Respondents Demographics}
Our survey sample included software engineers in diverse roles (see Table \ref{tab:roles}) with a median age of 32. Participants represented organisations from seven countries across three continents, spanning aviation, automotive, game design, infotainment, eLearning, cybersecurity, telecommunications, trade, and SE research and education domains. 




\begin{table}[!ht]
    \caption{Participants' areas and roles.}
    \begin{tabularx}{\linewidth}{llrr}
    \toprule
        Area & Roles & \# Participants &  Total\\
    \midrule
        & PhD Student            & 15 & \\
       \multirow{1}{*}{Academia}  & Researcher         & 7 & \multirow{1}{*}{27}\\
       & Professor              & 5  & \\
    \midrule
       &  Software Developer     & 10 & \\
      &    Software Engineer      & 6 & \\
       & Manager                & 5  & \\
     &   AI Engineer            & 4 & \\   
       \multirow{2}{*}{Industry} & Researcher             & 3 & \multirow{1}{*}{35}\\
       & Tech Lead              & 2  & \\
       & Software Designer      & 2  & \\
       & Software Tester        & 1  & \\
       & Application Specialist & 1  & \\
       & Applied Scientist      & 1 & \\
    \bottomrule
    \end{tabularx}
    \label{tab:roles}
\end{table}

\begin{figure}
    \centering
    \includegraphics[width=0.9\linewidth]{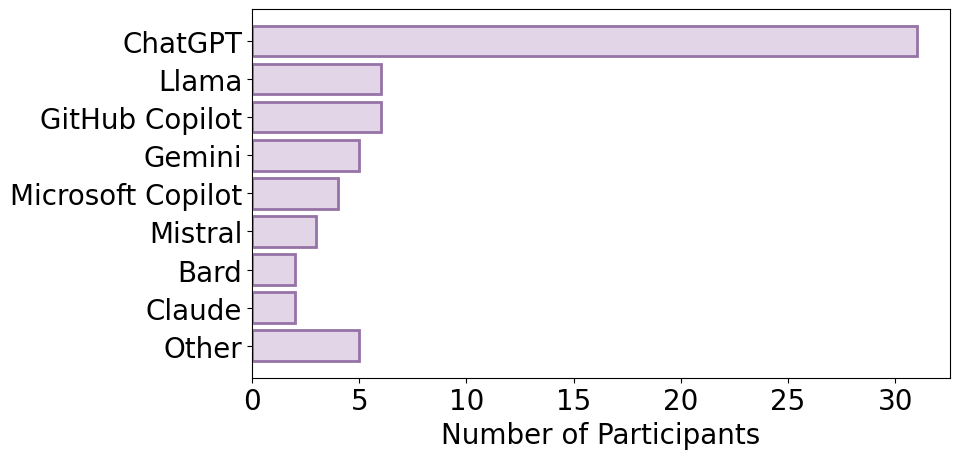}
    \caption{LLMs used by our participants.}
    \Description[LLMs]{LLMs used by our participants.}
    \label{fig:llms}
\end{figure}

Most participants (58 of 62) described themselves as `familiar" or ``very familiar" with LLMs. Figure \ref{fig:llms} shows the range of LLMs they use at work, with ChatGPT being the most popular.

\subsection{Usage of LLMs in Software Engineering Industry and Academia}
The results show that 56 participants (94.9\%) use LLMs occasionally, out of which 38 participants (66\%) use them on a weekly or daily basis. Only 3 (5.1\%) participants indicated that they rarely use LLMs at work.
Table \ref{tab:usages} shows the tasks for which academia and industry respondents apply them. We considered more than one answer per participant.

In \textbf{industry}, respondents use LLMs for \textbf{programming tasks} like code generation, debugging, and optimisation.
They also employ LLMs for \textbf{creative and communication tasks}, such as drafting emails and brainstorming ideas, and for generating and improving text. Additionally, LLMs help users on \textbf{learning new technologies and research} by providing starting points, best practices, and summaries of lengthy information.
Lastly, respondents view LLMs as digital assistants for \textbf{task management and problem-solving}, streamlining workflows and enhancing productivity. 

\begin{table}[!ht]
    \caption{LLM Usage Across Industrial and Academic Tasks}
    \begin{tabularx}{\linewidth}{Xlr}
    \toprule
        \textbf{Usage Group} & \textbf{Area} & \textbf{Total Count}  \\
    \midrule
        Writing Tasks                 &   Academia & 13  \\
        Programming Tasks              & Academia & 8   \\
        Educational Tasks              & Academia & 8  \\
        Research Related tasks         & Academia & 5   \\
        Non-Critical Tasks  & Academia & 5   \\
        \midrule
        Programming Tasks                & Industry & 18  \\
        Written Communication Tasks     & Industry & 10   \\
        Learning New Concepts & Industry & 8  \\
        Task Management & Industry & 5   \\
    \bottomrule
    \end{tabularx}
    \label{tab:usages}
\end{table}

In \textbf{academia}, LLMs are primarily used for \textbf{writing tasks}, including generating drafts, checking grammar, and providing content suggestions. Users find them helpful for managing busy work, such as email writing and idea generation, and for creating initial drafts for refinement. Additionally, participants view LLMs as \textbf{educational tools}, using them to understand new technologies and programming concepts or to assist in teaching.
For \textbf{programming tasks}, LLMs help write simple code, debug, and learn new coding concepts, offering initial code snippets and quick insights into technologies.

LLMs are also employed for \textbf{research-related tasks} such as summarising academic papers, generating ideas, and finding references. They assist in translating data, cleaning datasets, and extracting key information from research. 


\subsection{Emotions During LLM Interaction} 
Due to the complexity of the emotional responses to unexpected LLM interactions, participants often described multiple, layered feelings in their experiences. We mapped these feelings described to Willcox's emotions wheel (see Figure \ref{fig:wheel}) to assess on a more fine-grained level how evolved the emotions were. Following this framework, we could trace how initial feelings, such as anger (first-level emotion), might progress into more specific emotions, like annoyance (second level) and then frustration (third level), and quantify how often these emotions occurred at each stage.

\begin{figure}[!ht]
    \centering
    \includegraphics[width=0.9\linewidth]{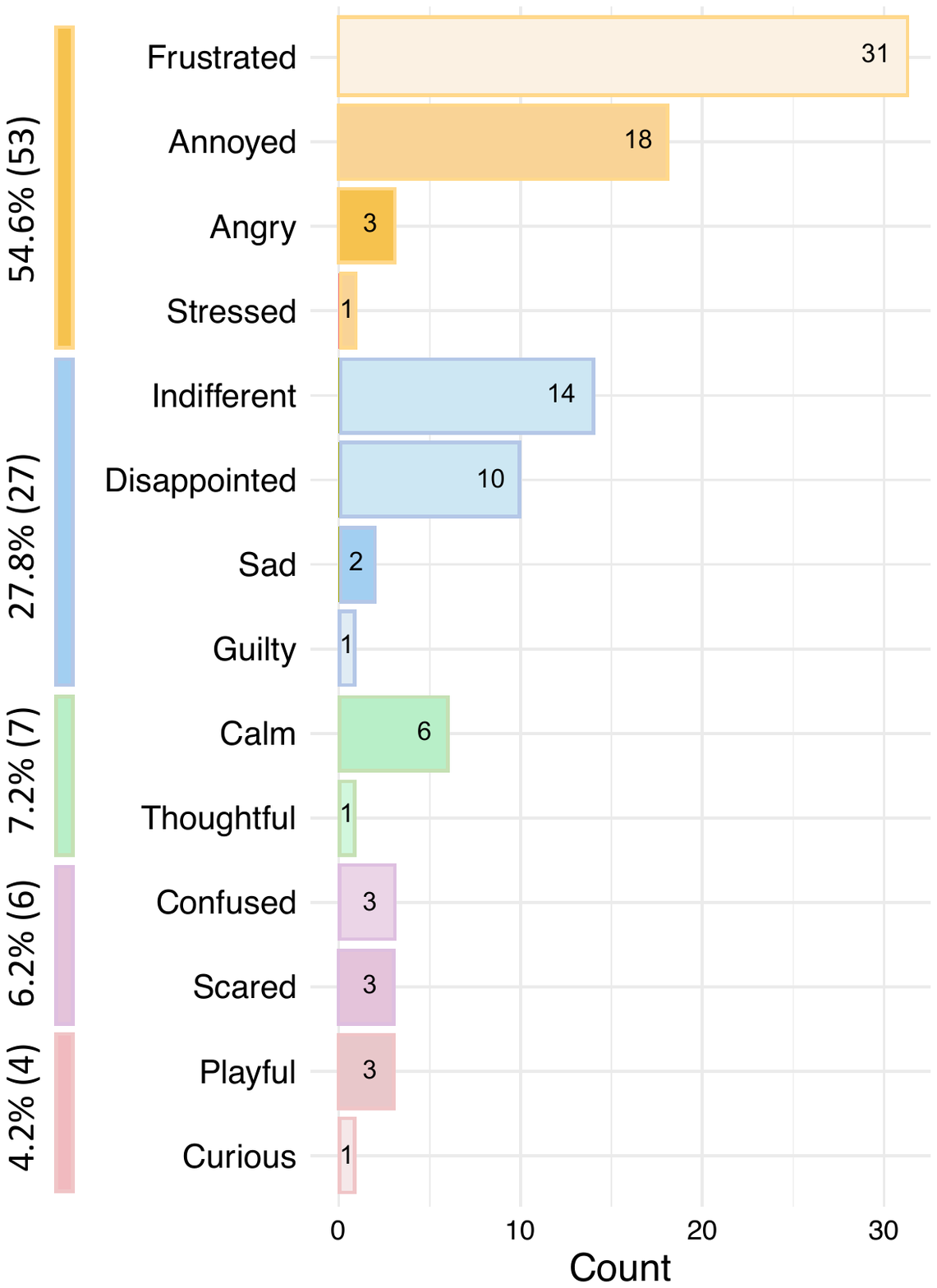}
    \caption{Emotional responses when receiving an incorrect answer. The colors map to Willcox’s emotions wheel in Figure \ref{fig:wheel}. }
    \Description[]{}
    \label{fig:emotions}
\end{figure}

In Figure \ref{fig:emotions}, we show the different emotions that were expressed by our participants for each category that map to Willcox's emotions wheel. For instance, we found that the most common emotions (54.6\%) are frustration or emotions that can develop into frustration, such as annoyance and anger. This poses potential challenges to well-being and smooth workflows, with a risk of cumulative emotional strain over time.

Many respondents (27.8\%) have also reported sadness-related emotions such as disappointment, indifference, or even guilt. In contrast to anger-related emotions, where respondents primarily blamed the LLM for its limitations, those who felt disappointed, sad, or guilty often turned the blame inward and criticised themselves for not being able to write the right prompt or meet their own expectations. When expectations were lower, the disappointment turned into indifference.

\dialoguegpt{quote}{
\pquote{29} ``Knowing how LLM[s] work, I typically have lower expectations. So I [don't] feel as frustrated or disappointed, particularly if I know that the task I asked is not trivial." }

Such expectations come from building knowledge about the LLM and understanding its capabilities and limitations based on previous interactions.

Less frequent reactions included positive emotions like calmness, thoughtfulness, playfulness, and curiosity, as well as negative emotions such as confusion and fear. This shows the varied and sometimes unexpected emotional responses that emerged. These emotions suggest that interacting with LLMs is not just a functional exchange but an exploratory experience for some software engineers. Curiosity, for example, has attached an investigative mindset often aligned with a trial-and-error approach. Playfulness, meanwhile, shows a willingness to engage with the LLM on a more open-ended basis. Fear introduces a new angle, suggesting that some engineers may feel a sense of responsibility if the interaction does not go as expected.

\subsection{Expectations When Interacting with LLMs}
\label{sec:expectations}

As shown in Table \ref{tab:expectations}, participants' expectations for LLMs extend beyond functionality to quality, usability, and versatility which are important factors in developing effective and trustworthy products in software development and design. Regarding \textit{quality} and \textit{performance}, engineers expect LLMs to consistently deliver correct and unhallucinated information without error or delay, as inconsistency can erode trust in the development workflow. 

\dialoguegpt{quote}{
\pquote{19} ``I rely on the LLM to provide accurate, relevant information that I can trust for both coding tasks and daily life management. It's like having an expert who gets it right the first time!''
}

In terms of \textit{understanding}, engineers value that LLMs understand context and intent, preferring models that ask clarification questions when necessary. They expect the LLM to understand the context without the need for a detailed context description in the initial prompt. This expectation is important since software research and development happen in dynamic and complex environments that require a lot of context such as best practices, policies, and relevant software artifacts. With the lack of consideration for such contexts, the outcome can become unusable and hard to integrate into the solution.


\dialoguegpt{quote}{
\pquote{19} ``I need an LLM that can seamlessly adapt to different contexts. Whether it's helping me with technical jargon, understanding project management lingo, [...], the LLM should be versatile enough to handle it all.''
}

Furthermore, since software engineering is a broad field with researchers from various sub-domains and specialties, researchers need to organize their texts and use language that fits the targeted audience. Therefore, software researchers both in academic and industrial organizations emphasized the importance of clarity and organisation of LLM-generated text, and that tailoring the answer to the \textit{structure} that the task needs are crucial.
For code-related tasks, engineers prefer responses starting with code snippets followed by explanations. When using LLMs to learn new concepts or explain artefacts, participants preferred elaborative answers. 

Additionally, they expect LLMs to be transparent by providing the source of the information and confidence estimation of the output accuracy to ensure \textit{information integrity} and support decision-making throughout the development lifecycle. This should also come with a need to protect the shared information in the chat. 

Finally, in terms of \textit{versatility}, engineers in industry increasingly expect LLMs to integrate with other tools and adapt to diverse workflows, reflecting the growing need for flexibility in software engineering environments. This also requires a \textit{usable} LLM, ideally with a user-friendly interface that enables intuitive interactions and a seamless integration of the LLM in the software-related tasks.


\begin{table}[ht]
    \caption{Users' Expectations when Using LLMs}
    \begin{tabularx}{\linewidth}{ll}
    \toprule
      \textbf{Themes} & \textbf{Categories} \\
    \midrule
        Quality & Accurate and Correct (39) \\ 
        & Reliability and Consistency (13) \\
        & No Hallucination (4) \\  
\midrule
        
        & Conciseness vs. thoroughness (14) \\ 
        \multirow{1}{*}{Answer Structure}& Complete with Examples (7) \\ 
        & Organised and Good Grammar (6) \\ 
\midrule
                
        Performance & Efficiency (response time) (20) \\ 
\midrule
        
         & Transparency (10) \\ 
        \multirow{1}{*}{Information Integrity}& Up-to-Date Information (4) \\ 
        & Data Security and Confidentiality (5) \\ 
\midrule
        
        Understanding & Intent Understanding (10) \\ 
        & Domain/Contextual Understanding (2) \\ 
\midrule
                
        Usability & Ease of Use (5) \\ 
\midrule
                
        Versatility & Integration with Other Tools (1) \\ 
        & Adaptability to Workflow (1) \\ 
        & Adaptability to Communication (2) \\ 
        \bottomrule
        
    \end{tabularx}
    \label{tab:expectations}
    
\end{table}

\subsection{Frustration Triggers in Software Engineering}
\label{sec:frustrations}
After exploring emotions in general related to receiving unexpected answers from LLMs, we focus on frustration-related emotions. We asked participants to describe specific situations where they felt frustrated when interacting with LLMs. From this, we identified patterns of triggers that cause LLMs to fail to meet engineers' expectations, leading to strains such as frustration. We outline the frustration triggers below.

\textbf{Repeated inaccuracies and hallucinations: } One of the most common causes of frustration was receiving repeated incorrect or hallucinated responses from the LLM. 

The definition of incorrectness varied among participants. Some examples of incorrect answers were uncompilable or buggy code, incorrect explanations of error messages, or incorrect factual information that was verified using other sources (e.g., an expert or a search engine).
Hallucination was described as nonsense explanations, references that do not exist, made-up packages, and invalid syntax in a programming language. 

Participants explicitly described that frustration arose when these issues persisted despite attempts to rephrase prompts or correct the LLM. For instance, participants referred to such situations as "annoying" or "disappointing" initially. However, they noted that \textit{repeated failures} led to frustration, describing the LLM as ``stubborn'' and ``insisting on an incorrect or hallucinated answer''.

\dialoguegpt{quote}{
\pquote{3} ``After several corrections, and repeating the prompt in different manners, it decided to reiterate the same wrong response."
}

This pattern of repeated failures can disrupt workflows in software engineering, where development cycles are often fast-paced and agile which requires more reliable and stable tools. For example, when the LLM provides a code that imports hallucinated libraries renders the code unusable which leads to wasted time fixing, debugging, or rewriting the entire implementation.

\dialoguegpt{quote}{
\pquote{38} ``During a coding problem, I was looking for the usage of a specific function in a library. I was frustrated when it provided a different [function] (which did not work or even exist)."
}

\textbf{Intent not understood: } 
Frustrations (or related emotions) are also caused when the software engineers feel that the LLM did not understand their prompt. Not understanding can be reflected in the response that is irrelevant to the initial question. Intent understanding was a common frustration trigger among our participants from industry and academia since in practice, engineers' queries are often highly technical and domain-specific and deal with complex software artefacts. Similarly, researchers and academics deal with novel techniques and niche problems that may cause the LLM to misunderstand the intent. Note that such misunderstandings are more common in general-purpose LLMs than in fine-tuned and specialized LLMs.


\dialoguegpt{quote}{
\pquote{29} ``I was trying to ask [LLM] to fill one specific cell in a notebook based on the others but it kept returning the same generic code for two cells instead of one. I had to talk to [LLM] like a child and say don’t do that and do this, and only this."
}

On another note, some participants pointed out their perceived usefulness of prompt programming and carefully constructing a prompt that would minimise such misunderstandings that can be caused by poor phrasing or the lack of context in the prompt. 

\dialoguegpt{quote}{
\pquote{5} ``I provide short and maybe unclear prompts [then] I usually get irrelevant responses. The better the prompt, the better response.''
}

While prompt programming has shown a high potential in enhancing the LLM outcome, it remains unclear whether it is effective in software engineering-related tasks.

\textbf{Personal preferences unmet: }
Many of our participants pointed out that they get frustrated with reasons related to their personal preferences. Some aspects of certain LLMs (e.g. answer structure) can be annoying to some engineers, and when the frequency of interactions with the LLM increases, the annoyance turns into frustration.
For example, two participants pointed out that an LLM apologising every time they tried to correct the LLM was a source of frustration since it can disrupt the workflow, especially during a refactoring or debugging process with the LLM, which results in a long conversation with many follow-up prompts. Others were frustrated with how the LLM they use only provides answers that are long, or only in bullet points. Forcing the LLM to structure an answer that aligns with their preferences required specifying many requirements and constraints in the prompt.

\dialoguegpt{quote}{
\pquote{35} ``[I get frustrated] when [LLM] gives too long answers. I quite often ask things that can be answered with a short sentence, but still I get half a page of ramblings back."
}

Such preferences depend on the task that the software engineer is solving. For example, important details when debugging might be hidden in long responses, while overly concise bullet points might omit crucial information needed to understand a system's architecture.

\textbf{Limitations of the LLM: }
LLM limitations (e.g., inability to perform specific tasks) or configuration constraints (e.g., context window size) are frustration triggers for software engineers. When an engineer attempts to force the LLM to overcome these limitations by prompting, it often leads to more frustration. For example, trying to generate a large application code in one prompt can result in missing lines and errors when the LLM hits its maximum token limit \citep{huggingface-llm}. The participants highlighted that there are some of the many tasks in software engineering research and practice that LLMs are just ``not good at''.

\dialoguegpt{quote}{
\pquote{56} ``[I got frustrated when] formatting of a table in latex, [had to] move to another LLM"}

\begin{figure}[!ht]
    \centering
    \includegraphics[width=\linewidth]{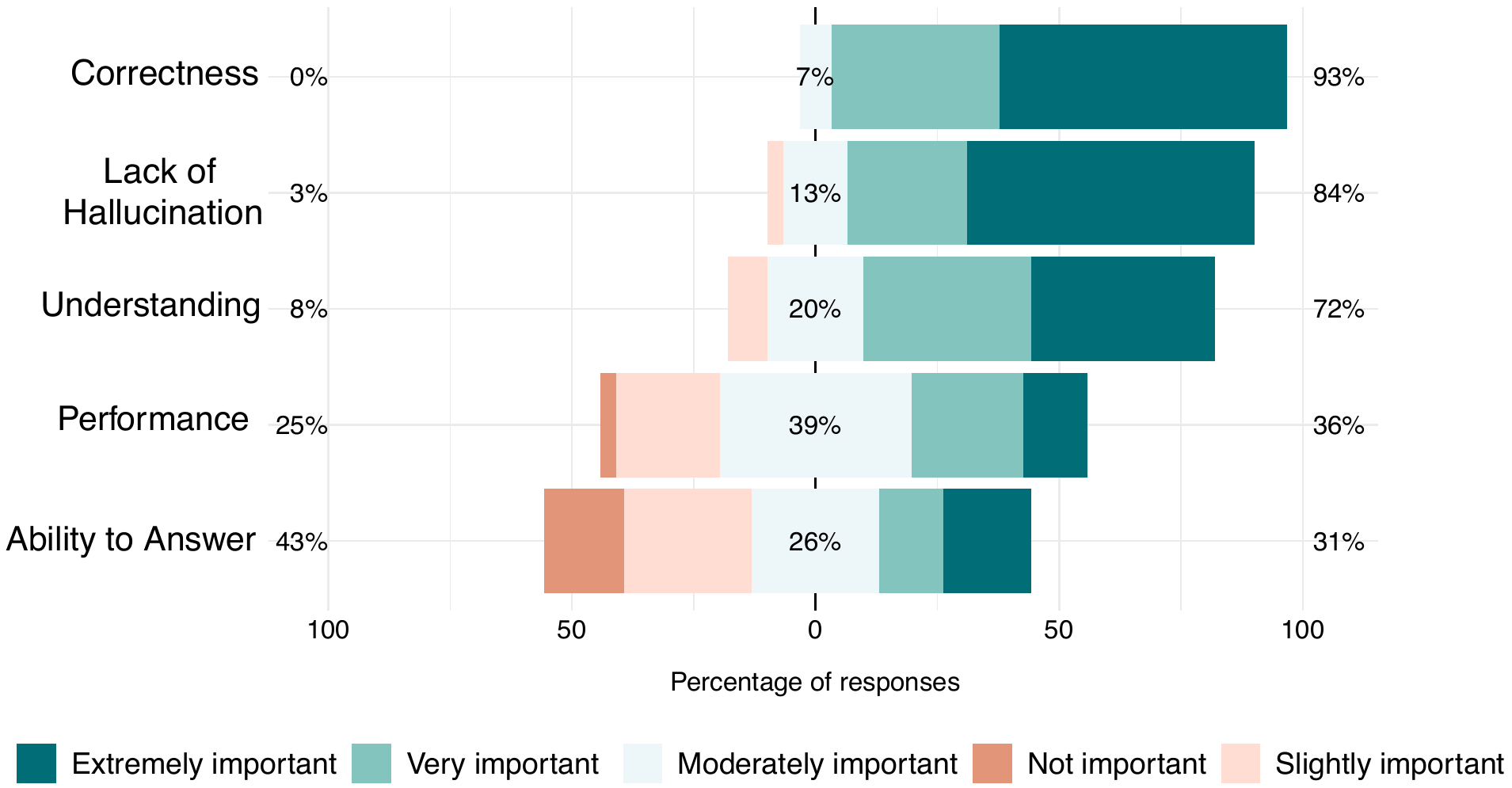}
    \caption{Likert-scale results of the importance level of different aspects of the LLMs that can impact the user experience. The scale ranges from Not Important (left - red) to Extremely Important (right - green). }
        \Description[]{}
    \label{fig:importance}
\end{figure}

Finally, as a verification question, we asked the participants to rate aspects of LLMs (correctness, lack of hallucinations, understanding, performance, and ability to answer) on a 5-point Likert scale based on how important they are in an LLM in order to ensure a better user experience.
The results in Figure \ref{fig:importance} resonate with the previous results where correctness, lack of hallucination, and correctness being the most important. In comparison, performance (e.g., response time) and ability to answer were seen as less critical.

\subsection{Unmet Expectations' Impact on Motivation}
\label{sec:motivation}

When the LLM failed to assist our participants, the participants' motivation to complete their task was influenced mainly in three ways. 

21.3\% (13 out of 62) of our respondents reported that their 
 \textbf{motivation decreased} when LLMs did not give the correct answer. Responses expressed frustration, stress, or disappointment, impacting participants' willingness to continue.
 Other participants commented that they eventually gave up. These respondents mentioned that after some effort, they decided to abandon the LLM and move on to other methods or stop altogether.
 \dialoguegpt{quote}{
\pquote{29} ``Eventually I give up [on the task], or report negative experimental results.''
}

Another group was formed of 77\% (48 out of 62) of the participants who were unfazed by the LLM's failure, treating it as a non-critical tool and continuing with the task with their \textbf{motivation not being affected}.

\dialoguegpt{quote}{
\pquote{8} ``My motivation is not affected, I just realise that the task will take longer''
}

In an interesting case, one participant mentioned that their
\textbf{motivation increased} which was due perceiving the interaction as a learning opportunity.

\dialoguegpt{quote}{\pquote{16} ``I usually understand the problem a lot more so I want to complete the task more''}

\subsection{User Actions for Improving LLM Interactions}

\begin{figure}[!ht]
    \centering
    \includegraphics[width=0.9\linewidth]{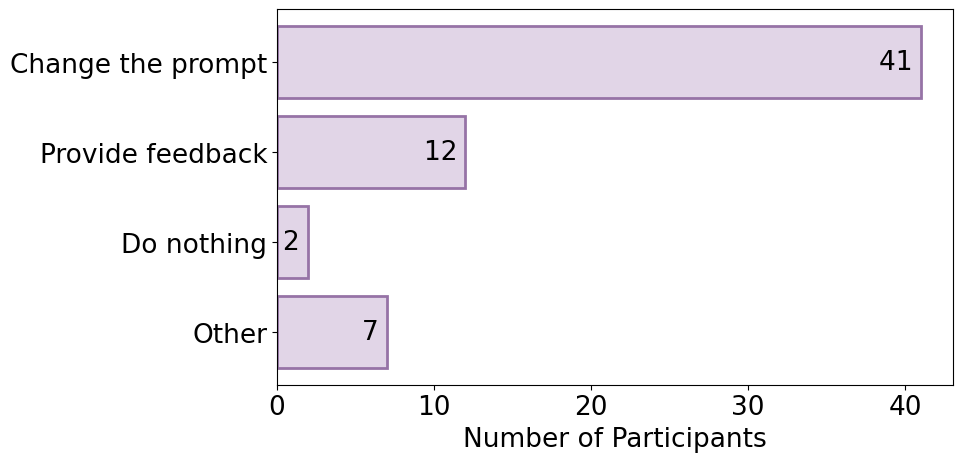}
    \caption{Actions done after receiving an unexpected answer from LLMs}
    \label{fig:Q9}
\end{figure}

When asked what actions participants typically take after receiving an unexpected answer from an LLM, the majority (41 out of 62 participants) said they changed the prompt to try again. A smaller group (12 participants) reported providing feedback to the LLM, while a few (2 participants) said they did nothing (see Figure \ref{fig:Q9}). In the ``Other'' category, participants described various strategies. Some combine multiple actions, such as changing the prompt, switching to another LLM, or even reverting to traditional search methods like Google or Wikipedia. A few participants mentioned disengaging from the LLM entirely or constructing the solution themselves.

\dialoguegpt{quote}{
\pquote{39} ``Sometimes the LLM hallucinates and puts me in loops. When I realize this, I resolve it myself using my human knowledge and [X] years of experience in the development field"
}

We asked our participants about the improvements they would like to see that would enhance their experience when using LLMs. They identified several areas that we grouped and sorted in Table \ref{tab:improvements} based on the number of times they were mentioned. The suggestions mainly align with the engineers' expectations (See Table \ref{tab:expectations}), though with a variation in the emphasis and distribution. 

Regarding the \textit{quality} of the outcome, the participants emphasised the importance of reducing hallucinations and improving response accuracy. They also commented on the need for enhanced analytical capacity to tackle more complex problems. However, the participants highlighted that while they expect an LLM to be accurate, transparency is more crucial for a less frustrating user experience. 
The transparency was described by the participants concerned \textbf{information integrity}, particularly specifying the data sources and the confidence level of the LLM's responses. This also includes stating any assumptions and reasonings by the LLM before providing an answer. This helps practitioners decide whether to rely on the LLM's suggestions, trust their own judgment, or seek assistance from a colleague instead. Establishing trust in this context is crucial, as it determines the main flow of the interaction, and hence the how the LLM output will be eventually used by the software engineer \cite{khojah2024conversations}.

Regarding the \textit{answer structure}, participants preferred concise and short answers (unless they were prompted otherwise). 
While in terms of \textit{understanding}, prompt comprehension was considered essential, with users expecting LLMs to ask clarifying questions when necessary. However, some participants mentioned that this could be a \textit{human-related} improvement where trainings that are designed specifically to learn to ``talk'' with the LLM are important to make communication more effective.


Finally, \textit{versatility} was another improvement that was consistently mentioned among our participants, they indicated the importance of handling more complex tasks and adapting to different user needs. The increased functionality to support a broader range of tasks was also crucial for improving overall user satisfaction. This is especially critical as software engineers researchers and practitioners often work with a variety of tasks that range from learning new concepts and tasks to implementing products and using tools.
Furthermore, practitioners particularly stressed the importance LLM \textit{performance} in terms of increased memory and learning capabilities, allowing LLMs to retain relevant information and build on past interactions with a faster processing time.

Overall, these requirements guide chatbot designers to understand practitioners' and researchers' needs and priorities in software engineering. For instance, the next-generation LLMs may need to direct efforts toward transparency rather than performance when aiming for a better user experience.

\begin{table}[ht]
    \caption{Improvements for Better User Experience}
    \begin{tabularx}{\linewidth}{ll}
    \toprule
      \textbf{Themes} & \textbf{Categories} \\ 
      \midrule
      
        & Transparency (source, confidence) (16) \\
        \multirow{1}{*}{Information Integrity}& Trust and Data Security (4) \\
        & Relevant information (Up-to-date) (2) \\
        & State reasoning and assumptions (2) \\
        \midrule

        Versatility & Adaptability in communication (11) \\ 
        & Integration in project environment (2) \\
        & Extended Functionality (2) \\ 
        \midrule
        
        Quality & Improved Response Accuracy (6) \\ 
        & Reduced Hallucinations (4) \\ 
        & Less creative (3) \\ 
        \midrule

        Understanding & Context understanding (5) \\ 
        & Clarification questions (3) \\ 
        & Intent Understanding (2) \\ 
        \midrule
        
        \multirow{1}{*}{Answer Structure} & Elaborative answers on-demand (4) \\ 
        & Consistency in Responses (3) \\ 
        \midrule
        
        Performance & Higher memory utilisation (4) \\ 
        & Efficiency (processing time) (1) \\ 
        \midrule
        
        Human-related & Training for engineers (2) \\ 
        \bottomrule
        
    \end{tabularx}
    \label{tab:improvements}
\end{table}

\section{Discussion}
In this section, we answer our research questions highlighting the main takeaway per question.

\subsection{Frustrations in the Context of LLM Interaction}

We identified four main frustration triggers for software engineers using LLMs—accuracy issues, hallucinations, misunderstandings, unmet preferences, and LLMs limitations— pointing to significant challenges with potential long-term repercussions. For example, repeatedly dealing with inaccurate, buggy or non-standard code disrupts workflow, forcing engineers to spend extra time debugging \citep{fitzgerald2008debugging} or reworking tasks that an efficient tool should minimise. This is particularly important if the code produced by the LLM is less maintainable \citep{liu2024refining} and can be more frustrating than writing the code with no LLM assistance \citep{weisz2022better}. 
This extra workload impacts project timelines and creates a deeper frustration that could enforce frustration and emotional strain \citep{ceaparu2004determining,lazar2006workplace} as engineers may feel that they have spent longer than planned to reach their goal (e.g., completing a task).
Additionally, participants' frustration resulted from other emotions, including anger and annoyance (see Figure 4).

On Wilcox's emotion wheel, these frustrations can be linked to various emotions within the frustration spectrum, including anger, annoyance, and confusion. For instance, frustration over inaccurate or faulty code can easily evolve into anger when engineers feel a lack of control over the situation, especially if the tool is supposed to improve efficiency. Similarly, when an LLM produces outputs that deviate from expectations, it can lead to another emotion within the spectrum, for example, annoyance, particularly when the tool fails to meet personal preferences or the engineer's standards. In addition, misunderstandings or incorrect outputs might lead to confusion as engineers try to reconcile the LLM's output with their original intent.

Studies on GitHub Copilot \citep{usability2023frustration, usability2023frustration} showed similar emotions, especially around data privacy concerns, intrusive code suggestions and usability, particularly unnecessary large code suggestions. Eshraghian at al. \citep{eshraghian2024ai} explained that frustration and anger can come from feeling a threat without being able to control it. In the previous example, the threat of leaking confidential data with very little control over it (i.e., to use the LLM, they need to accept the policy) was the trigger for frustration.

Unlike other domains where frustration often stems from performance issues (e.g., system crashes) \citep{lazar2006workplace}, our participants did not report such frustrations, likely due to recent LLMs being stable and fast (Table \ref{tab:expectations}). Other frustrations in the medical domain arise from the emotionally exhausting work environment along with their dependence on the technology (e.g., to document patient data) \citep{opoku2015user, tawfik2021frustration}. In contrast, software engineers can still rely on their expertise or alternative tools, as LLMs are not essential for task completion (Section \ref{sec:motivation}).



\cristysbox{
    \ta{Takeaway:}
    Frustration triggers studied in software engineer literature (including ours) come from spending extra time refining output, unlike in existing studies in other domains where it is due to performance and usability issues.
}



\subsection{Impact of Frustrating Experiences on Motivation}

While most of our participants expressed frustration (or similar emotions), their motivation to complete the task was not necessarily affected. This can suggest that although frustration may reside in the emotion wheel's 'anger' or 'irritation' sections, the engineers' resilience and coping mechanisms allowed them to manage these emotions without diminishing motivation.

Our results showed that participants often felt demotivated when an LLM failed to meet their expectations, such as when it did not assist them as intended, frequently leading to frustration. However, most of those who felt frustrated reported that their motivation remained intact, likely due to perceiving the LLM as merely one tool among others in achieving their goal. When an LLM could not provide the necessary assistance, participants commonly resorted to alternative solutions, such as using search engines like Google or relying on their expertise to complete the task independently. These observations align with findings by Franca et al. \citep{francca2014motivated}, who explored the connection between motivation and the satisfaction and happiness of software engineers, revealing that happiness slightly overlaps with but does not correlate with motivation. 

These findings suggest that, despite facing emotional challenges, software engineers maintain their motivation to persist with demanding tasks.
Although frustration may not directly impact motivation, it remains a critical factor to consider due to its known connection with burnout. Sustained frustration is still significant as it contributes to emotional strain \citep{tawfik2021frustration} —a known precursor to burnout in high-demand professions like SE. Thus, understanding and managing frustration, even when the motivation appears unaffected, is essential in supporting the long-term well-being of software engineers.

\cristysbox{
    \ta{Takeaway:}Although frustration occurs when LLMs fail to meet expectations, it generally does not diminish the motivation to complete tasks.
}

\subsection{Towards a Less-Frustrating User Experience}

When designing chatbots and LLMs, it is essential to prioritise not only high accuracy and performance but also emotional intelligence such as recognising user frustration. Wilcox' emotion wheel provides a nuanced view of emotional states, categorising emotions into primary and secondary feelings. Recognising and responding to these emotions in real-time is key. 

The emotional intelligence of LLMs has been explored by Wang et al. \citep{wang2023emotional} where they saw that newer-generation LLMs such as GPT-4 (at the time of the study) show a better ability to understand user emotions that compared to humans' emotional intelligence.
However, recognising emotions is insufficient as the LLM should also know how to act accordingly. In Section \ref{sec:frustrations}, we saw that even minor LLM behaviours such as apologising after receiving feedback was making the engineers' experience more frustrating. This was also described by Erlenhov et al \citep{erlenhov2019current} about ideal development bots adapting communication to different individuals.
While several studies focus on enhancing LLMs, we recognise that these systems will always have room for improvement.  Therefore, our focus in this study shifts to the human element. We propose enabling users with the knowledge and skills to navigate LLMs effectively to reduce frustration and improve overall user experience.
The goal is to address frustration triggers that arise during their use. Specifically, triggers such as misunderstanding of the intent, unsatisfying personal preferences, or even getting incorrect answers can be minimised by prompt engineering the query before sending it to the LLM. Prompt engineering can involve incorporating prompt techniques (such as Few-shot learning), relevant contextual information (e.g., system description), or constraints about the output (e.g., the output structure). Other frustration triggers, such as hallucinations and limitations of LLMs, can be minimised if the engineers use the LLM according to its capabilities and limitations. Since unmet expectations are among the primary frustration triggers\citep{levine2022unhappy} (see Section \ref{sec:expectations}), providing software engineers with training on effective usage and clear understanding of LLM capabilities to set realistic expectations can help reduce disappointment and enhance user satisfaction. 

Additionally, raising awareness about potential frustration triggers is important; engineers can manage their reactions accordingly if they recognise the likelihood of frustration in certain situations. For instance, using coping strategies rather than repeatedly attempting to elicit a perfect answer from the LLM. Therefore, we suggest that software engineers need training on how to use LLMs safely. This idea was also discussed by Barman et al. \citep{barman2024beyond} where they propose providing guidelines for LLM users to know how to interact with different LLMs, for instance, if it is appropriate to generate artefacts or only to get some guidance. Our participants commented that they were familiar with LLMs; however, most reported frustration, raising questions about whether they truly understand how to leverage LLMs effectively. Familiarity does not necessarily equate to proficiency\citep{ladwig2012perceived}, stressing the need for improved training and guidance on optimal usage strategies. 
A complete understanding of LLM capabilities and limitations can help users to manage their expectations.

\cristysbox{
    \ta{Takeaway:}
A less frustrating experience arises from combining ``emotionally intelligent'' LLMs with engineers’ awareness to manage their expectations and reactions to frustration triggers. }

\subsection{Threats to Validity}

In this section, we explain the strategies to address this study's threats to validity.

\textbf{Internal Validity:}
To ensure internal validity, we considered several biases and employed mitigations accordingly. Self-selection bias: Participation in the survey was voluntary; hence, individuals with particularly strong positive or negative experiences with LLMs might be overrepresented, skewing the data. To mitigate this, we tried to recruit diverse participants across different experience levels, regions, and fields. To address social desirability bias, we collected anonymous data by including a statement at the beginning of the survey and avoiding personal questions. This approach aimed to prevent participants from feeling pressured to align their responses with what they perceived as socially or professionally acceptable. This could lead to underreporting frustration to appear more competent with new technologies.

\textbf{External Validity:}
Our sample size of software practitioners and academics can limit to extent to which our findings can be generalized to the broader population of software engineers, to minimize this threat while avoiding overrepresenting certain groups or regions, we targeted respondents from different countries and several fields within SE. Similarly, we employed stratified sampling to ensure a balanced representation across demographics, skill levels, and industries.

\textbf{Construct Validity:}
We operationalised key concepts like frustration and hallucination to guarantee construct validity, adding their definitions. Additionally, we provided examples throughout the survey to clarify the scenarios we were exploring. Finally, we explained the Likert scale by adding information on how to measure each level. Further, we piloted the survey to ensure clarity. We used the feedback to fix ambiguous questions, clarify terms, and minimise confusion.

\section{Conclusion}

This study focused on the emotional strains, particularly frustration, experienced by software engineers when interacting with LLMs and not being assisted as intended. By identifying main triggers, such as the correctness and reliability of responses and issues related to personalisation, we emphasise that understanding the emotional impacts of LLM use in SE is important. This study's insights bring attention to the potential risks to productivity and mental health if emotional responses go unaddressed. Future research should further explore the psychological implications of LLM use, focusing on sustainable strategies to support the well-being of software engineers and optimise their user experience with AI tools.

\bibliographystyle{ACM-Reference-Format}
\bibliography{bib.bib}

\end{document}